\documentclass[aps,prl,twocolumn,floatfix,superscriptaddress]{revtex4-1}

\usepackage{rotating}
\usepackage{epstopdf}
\usepackage{graphicx}
\usepackage{natbib}
\usepackage{epstopdf}
\usepackage{amsmath}
\usepackage{amssymb}
\usepackage{mathbbol}
\usepackage{xcolor}
\usepackage{lipsum}
\usepackage{amssymb}
\usepackage{mwe}
\usepackage{lipsum} 
\usepackage{siunitx}
\usepackage[normalem]{ulem}

\usepackage{empheq}
\usepackage{afterpage}
\usepackage[colorlinks,citecolor=blue,urlcolor=red]{hyperref}
\usepackage{cleveref}

\newcommand{\beq}{\begin{equation}}
\newcommand{\eeq}{\end{equation} \smallskip}
\newcommand{\beqy}{\begin{eqnarray}}
\newcommand{\eeqy}{\end{eqnarray} \smallskip}

\newcommand{\bit}{\begin{itemize}}
\newcommand{\eit}{\end{itemize}}
\newcommand{\bmat}{\begin{pmatrix}}
\newcommand{\emat}{\end{pmatrix}}

\newcommand{\blue}[1]{\textcolor{blue}{#1}}

\newcommand{\paolo}[1]{\textcolor{olive}{#1}}

\begin{document}

\title{Dynamics of Onsager vortex clustering in decaying turbulent polariton quantum fluids}

\author{P. Comaron}
\email[Corresponding author: ]{p.comaron@ucl.ac.uk} 
\address{Institute of Physics, Polish Academy of Sciences, Al. Lotników 32/46, 02-668 Warsaw, Poland}
\address{Department of Physics and Astronomy, University College London,
Gower Street, London, WC1E 6BT, United Kingdom}

\author{R. Panico}
\address{CNR NANOTEC, Institute of Nanotechnology, Via Monteroni, 73100 Lecce, Italy}

\author{D. Ballarini}
\address{CNR NANOTEC, Institute of Nanotechnology, Via Monteroni, 73100 Lecce, Italy}

\author{M. Matuszewski}
\address{Institute of Physics, Polish Academy of Sciences, Al. Lotników 32/46, 02-668 Warsaw, Poland}
\address{Center for Theoretical Physics, Polish Academy of Sciences, Al. Lotników 32/46, 02-668 Warsaw, Poland}

\begin{abstract}
We investigate the turbulent properties of a confined driven-dissipative polariton quantum fluid after a pulsed excitation.
Using numerical simulations, we provide insight into the vortex clustering processes that emerge during the relaxation dynamics of the initially injected vortex cloud.
A confrontation between conservative and non-conservative dynamics reveals that the onset of clusterization strongly depends on the interplay between the different characteristic system lengths and time-scales at stake, with an additional time-scale due to dissipation in the non-conservative case.
Quantification of the clustering observables allows us to numerically characterize the optimal conditions for observing Onsager condensation in decaying polariton systems, demonstrating its experimental reachability under pulse excitation. These findings hold significance for exploring the onset of turbulent dynamics in open systems, spanning both classical and quantum domains.

\end{abstract}

\maketitle

Quantum turbulence~\cite{barenghi_skrbek_sreenivasan_2023, bagnato2022}, characterized by the re-organization of chaotic motion of topological defects and waves in quantum fluids, has been widely explored in many systems, such as Bose Einstein condensates of ultracold atoms, superfluid helium He$^3$ and He$^4$~\cite{white2014,Barenghi_book, roati2021}.
In two dimensions (2D), nonequilibrium turbulent phenomena in quantum systems have sparked renewed interest over the past decade, both experimentally~\cite{chesler2013holographic,weiler2008spontaneous,dalibard-2D-KZ-2015,johnstone2019evolution,gauthier2019giant} and theoretically~\cite{kagan1997evolution,berloff2002scenario,kobayashi2005kolmogorov,kobayashi2007quantum,white2010,Bradley2012,muller2023}.
Photonic systems are emerging as an alternative platform for the investigation of 2D turbulence, allowing the direct measurement of the phase of the quantum fluid with interferometric techniques~\paolo{\cite{Drori2023, qfdispersion2020, BakerRasooli2023,Congy2024}}. In particular, the hybrid light-matter nature of exciton-polariton in semiconductor microcavities enables effective photon–photon interactions and a light mass, which lead to the observation of superfluidity~\cite{Carusotto2004,Amo2009,Lerario2017} and Bose–Einstein condensation \cite{kasprzak2006bose,Wertz2009}. While extensive investigations have been performed on the dynamics of quantum vortices~\cite{lagoudakis2008quantized, dominici2015},
which can be formed during the nonlinear dynamics of the polariton quantum fluid~\cite{maitre2021,Nardin2011,Sanvitto2011,Gnusov2023} or externally imprinted by optical means~\cite{caputo2019,Panico2021}, their clustering dynamics was addressed only recently~\cite{panico2023,conformal2023}.

A generalised picture of two dimensional confined quantum systems dynamics is given by means of statistical tools, similar to those employed in the study of classical chaos~\cite{kraichnan1967inertial}. 
In a pioneering paper, Onsager was able to justify the formation of vortex clusters ---directly connected with the inverse cascade of the system's energy from small to large length scales --- using only equilibrium arguments.
The theory predicts a transition from a configuration of randomly distributed vortices to more energetic ---and less-entropic--- configurations of giant clouds of defects, the so-called Onsager vortices~\cite{simula2014prl}. 
Previously investigated in numerous numerical works~\cite{Bradley2012,Reeves2013prl,Groszek2018,Tsubota_Onsager}, emergence of Onsager order has been reported in ultracold gases~\cite{gauthier2019giant,johnstone2019evolution}.
The spontaneous emergence of order from chaos, even in the absence of a constant energy injection (decaying turbulence), has been explained by means of an evaporative-heating mechanism: the number of vortices decreases, due to vortex-pairs annihilation and vortex losses at the edges of the trap, but the total kinetic energy of the system is conserved, leading to an increase of the mean energy per vortex~\cite{Billam2014,Kanai2021}. This leads the randomly distributed vortex cloud to eventually evolve to a negative temperature, low-entropic state. 

In the aforementioned studies, this phenomenon is discussed for systems where the particle number is conserved. Notably, the formation of vortex clusters is generally not guaranteed in a non-conservative system~\cite{Groszek2020}, where the decaying energy is expected to preclude the relaxation towards negative-temperature states~\cite{Groszek2018}.
It is therefore relevant to investigate the conditions required for the emergence of Onsager clustering in a paradigmatic driven-dissipative system such as exciton-polariton quantum fluids.
The question of i)~whether a decaying vortex cloud in a polariton condensate formed after an initial pulsed excitation can still exhibit clustering of same-sign vortices toward Onsager order, and ii) what are the underlying competing mechanisms which regulate their dynamics, and their importance,
remain therefore unanswered and of high relevance for a comprehensive understanding of turbulent regimes in light quantum fluids.

In this paper, we numerically investigate and extract the optimal conditions for the observation of vortex clustering in the decaying turbulence regime of a dissipative polariton condensate.
We demonstrate that the onset of order depends on the interplay of different spatial and temporal scales, and it is eventually the result of the competition between opposite tendencies driven by particle dissipation and vortex interactions, respectively.
Moreover, we discuss the interaction between vortices and sound waves, showing the important role played by the initial density of the polariton fluid.
Finally, we show that the observation of Onsager condensation in decaying polariton fluids is experimentally achievable within realistic length and time scales. 

The document is structured as follows. In the first section of the paper, we describe our numerical modeling and the parameters used.
In the second section, we discuss how we model and probe the vortex dynamics after a high-entropy initial configuration of vortices is imprinted in the polariton condensate. 
In section III, we investigate the conservative polariton case, revealing how vortex clustering is affected by the intervortex distance.
In section IV, we investigate the vortex dynamics in dissipative systems, using different decay rates.
In section V, we discuss how clustering dynamics changes in a dissipative system as a function of both the initial particle density and the intervortex distance.
Finally, we summarise our conclusions in the last section.

\paragraph{\blue{Section I - Theoretical model and system parameters.}}
The dynamics of the polariton fluid is described by means of a generalized equation of motion for the 2D polariton field $\psi$ for the lower-polariton branch, a function of the position $\mathrm{\textbf{r}}=(x,y)$ and time $t$;
it can be derived from both truncated Wigner (TW) and Keldysh field theories~\cite{carusotto2013quantum,szymanska2006nonequilibrium} and reads ($\hbar=1$):
\begin{equation}
\hspace{-4mm}id \psi = dt \bigg[ - \frac{
	\nabla^2}{2 m } + g|{\psi}|^2_{-} + V(\textbf{r}) - \frac{i}{2} \gamma_\mathrm{LP} \\ \bigg]
\psi +  dW.
\label{eq:SGPE_pol}
\end{equation}
Here $m$ represents the polariton mass, {$g$} the polariton-polariton interaction strength, $\gamma_\mathrm{LP}$ the polariton loss rate $\gamma_\mathrm{LP} = 1/\tau_\mathrm{LP}$, with $\tau_\mathrm{LP}$ the polariton lifetime.
We consider a simplified case where the complex relaxation processes~\cite{woutersLiew2010,comaron2018dynamical} are disregarded, and where the exciton reservoir is treated as adiabatically following the condensate dynamics~\cite{bobrovska2015}. 

The renormalized density $|{\psi}|^2_{-} \equiv \left|{\psi} \right|^2 - {1}/{2dV}$  {includes} the subtraction of the Wigner commutator contribution (where $dV=a^2$ is the {volume element} of {our $2D$ grid of} spacing $a$). 
The zero-mean white Wiener noise $dW$ fulfills
\begin{eqnarray}
\left <dW(\textbf{r},t)dW(\textbf{r}^\prime,t)\right> &=& 0, \nonumber \\
\left < d W^*(\textbf{r},t)dW(\textbf{r}^\prime,t)\right> &=& (\gamma_\mathrm{LP}/2) \delta_{\textbf{r},\textbf{r}^\prime}dt,
\label{eq:corr_noise}
\end{eqnarray}
as required by the TW approach.
The two-dimensional fluid of light is confined in a radially-symmetric potential $V(\textbf{r})$ with diameter $D$.
In typical experiments, the potential landscape can be created by means of an additional laser that confines the polariton fluid or it can be obtained by lithographic etching of the sample surface. Both optical and lithographic techniques enable further engineering of the polariton dispersion, an additional tool we anticipate to be useful to control the dynamics of polariton quantum fluid in future investigations.
For simplicity, in this work, the potential is modeled as a hard-bounded box. 
In order to avoid numerical boundary effects, we choose values of the ratio $\alpha = D/L$ (between the diameter $D$ and the extent of the numerical grid $L$) less than $0.5$.
Typical experimental parameters for the polariton system are used~\cite{Panico2021,panico2023}: 
$\tau_\mathrm{LP} = 0.2 \mathrm{ns}$, $m = 0.22 \ \si{ps^2 meV \mu m^{-2}} = 3.52 \times 10^{-35} \ \si{Kg}$, $g = 5 \times 10^{-3} \ \si{meV \mu m^2}$.
We solve Eq.~\eqref{eq:SGPE_pol} using an eigth-order Runge-Kutta method on a $512^2$ numerical grid.

\begin{figure}
	\centering
    \includegraphics[width=\columnwidth]{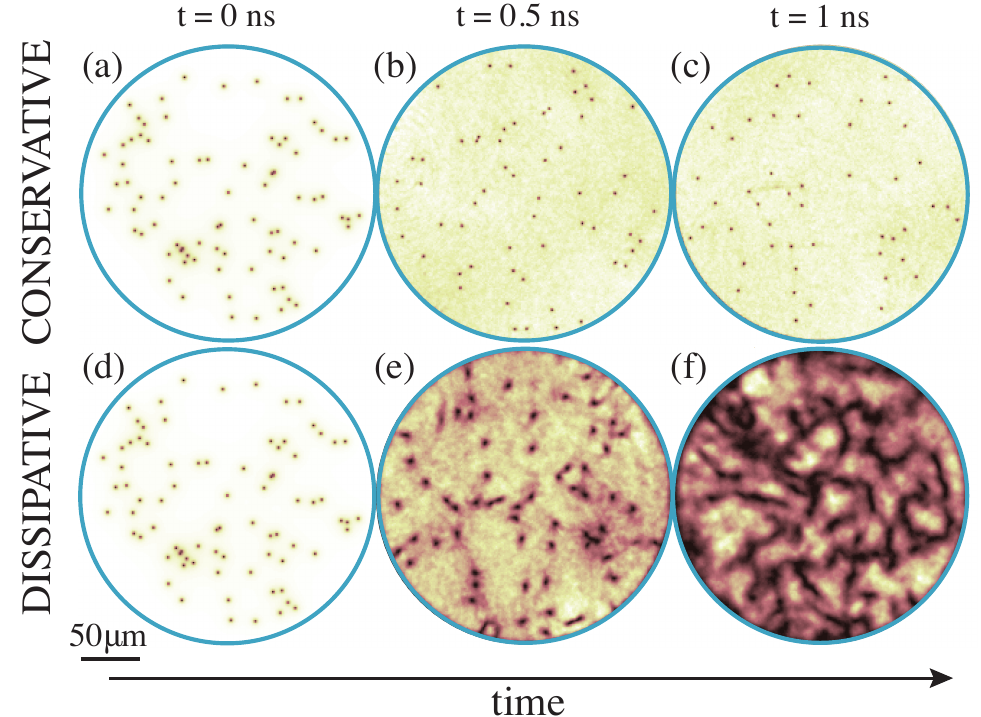}
	\caption{\textbf{Conservative and dissipative temporal evolution}.
    The vortex gas dynamics is depicted for the conservative \textbf{(a, b, c)} and dissipative \textbf{(d, e, f)} cases. Snapshots are taken at a time ${t=0\mathrm{ns}}$ (a, d), ${t= 0.5 \mathrm{ns}}$ (central panels) and ${t= 1 \mathrm{ns}}$ (right panels). 
    While in the conservative dynamics, the polariton density and vortex healing length are constant in time, in the dissipative case the density decays while the vortex healing length increases, modifying the evaporative heating mechanism.
    The top row illustrates the conservative case {with an initial healing length ${\xi_0 = 0.7\mathrm{\mu m}}$ {(corresponding to ${n_0=400 \mathrm{\mu m^{-2}}}$)} and initial inter-vortex length ${\ell_\mathrm{v}^0 = 28 \mathrm{\mu m}}$ (corresponding to ${D=250 \mathrm{\mu m}}$)}, respective to the pink solid line in Fig.~\ref{fig2}. The bottom row shows the dissipative case, respective to the green solid lines in Figs.~\ref{fig3},\ref{fig4}, and to the green point in Fig.~\ref{fig5}(a). 
}
	\label{fig1}
\end{figure}

%
\paragraph{\blue{Section II: Modelling the vortex dynamics.}}
In this study, we simulate turbulent processes of a decaying driven-dissipative system after an initial pulsed excitation.
To do so, we solve the system dynamics starting from a condensate at $t = 0$ ns with homogeneous initial density $n_0 = |\psi|^2(t=0)$.
The spatially homogeneous but arbitrary phase emulates a quasi-resonant laser excitation.
This initial state can be also thought of as the density state corresponding to a non-resonant pumping with strength $P(t=0) = \gamma_\mathrm{LP}(n_0/n_\mathrm{s}+1)$, where $n_\mathrm{s}$ is the system saturation density~\cite{wouters2007}.
At the same time ($t=0 \ \mathrm{ns}$), a random configuration of $N_\mathrm{v}^{0} = 40 \mathrm{(vortices)} + 40 \mathrm{(anti-vortices)}$ is imprinted; this configuration corresponds to an entropy-dominated vortex gas state~\cite{simula2014prl}.
{The position of the $i$-th vortex, centered at $\textbf{r}_i = (x_0-x_i,y_0-y_i)$, with $\textbf{r}_c =(x_0,y_0)$ the center of the potential, is randomly chosen at each stochastic realization.
To imprint the vortex cloud, the polariton wavefunction is initialized as $\psi(\textbf{r}) = \sqrt{n_0} \ \Pi_i [\textbf{r}_i/\sqrt{(\xi_0/\xi_c)^2 + \textbf{r}_i^2} \exp{(\pm i \theta_i)}]$, with $\xi_c$ a parameter regulating the shape of the vortex core.
}
We then let the system evolve with Eqs.~\eqref{eq:SGPE_pol}-\eqref{eq:corr_noise}.
Topological defects are numerically detected at each snapshot of the evolution as in Ref.~\cite{comaron2018dynamical}.
Vortex number and their positions are inferred from the phase gradients around closed paths of each grid point after a Gaussian filtering is applied in order to get rid of highest-order fluctuations~{(see Ref.~\cite{Comaron2021} for details on the vortex tracking algorithm)}.
Evaluation of the position, vorticity, and charge of each vortex allows us to track the behavior of the vortex gas, and to extract the relevant clustering observables. 

In order to quantify the clustering processes, we monitor the evolution of the averaged dipole moment $ d_\mathrm{m}  =  |\textbf{d}_\mathrm{m}|  = |\sum^N_{i=1} \kappa_i \textbf{r}_i|$ and the correlation function $C = \frac{1}{N}\sum_{i=1}^{N}c_i$~\cite{gauthier2019giant}. Here $N$ is the total vortex number, $\kappa_i$ the circulation sign, $\textbf{r}_i$ the position, and $c_i$ the product of the circulations between the $i$-th vortex and its nearest neighbor.
As such, increasing values of C correspond to higher energetic states of the vortex gas. 
In the Onsager model, negative temperatures are associated to $C>0$, whereas $C=0$ corresponds to the infinite temperature limit (maximum entropy) and $C=-1$ is the lowest (positive) temperature~\cite{kraichnan1967inertial,simula2014prl}. The limits $C={-1,0,1}$, correspond to a gas of vortex-antivortex pairs, a completely random configuration, and to configurations where same-sign vortices gather into giant clusters, respectively~\cite{Groszek2018}. 
We also track the evolution of the average distance between the vortex cores  $\ell_\mathrm{v}$. In a disc with diameter $D$, by modelling the single vortices as rigid objects with radius $\ell_\mathrm{v}/2$, the average inter-vortex length can be calculated as $\ell_\mathrm{v}= {D  N_\mathrm{v}}^{-1/2}$.

In the following analysis, we show how the nature of the clustering processes strongly depends on the interplay of  different system lengths and  timescales of the physical processes, namely the healing length (proportional to the size of the vortex core)
\begin{equation}
    \xi(t) = {1}/{\sqrt{2 m g |{\psi(t)}|^2}},
    \label{eq:heal_length}
\end{equation}
the circular  trapping diameter $D$ and the average inter-vortex length
$\ell_\mathrm{v}(t)$,
where $\xi(t) < \ell_\mathrm{v}(t) < D$.
Here, the inter-vortex length is defined as $\ell_\mathrm{v}(t)= {D}{ \langle N_\mathrm{v}(t) \rangle}^{-1/2}$. 
In our numerical experiments, we impose the initial polariton density $n_0$ ---which also sets the initial healing length of the system $\xi_0$--- as well as the number of initial vortices $N_\mathrm{v}^0$.
Fixing $N_\mathrm{v}^0$ allows us to use the disc diameter $D$ as a parameter controlling the initial intra-vortex length $\ell_\mathrm{v}^0 = D/\sqrt{N_\mathrm{v}^0}$.
We are careful to set such a ratio to be about a decade, large enough to avoid finite-size effects. 
We chose a configuration of vortices that is highly random and free of clusters, by carefully checking that the initial condition for the dipole moment, averaged over the $\mathcal{N}$ realizations, always reads $\left< d_\mathrm{m} \right>_\mathcal{N}(t=0)<3 \mathrm{\mu m}$.
Since we are interested in the early stages of the dynamics, in all our simulations we probe the polaritons up to $1$ ns (5 times the value of the polariton lifetime in typical high-quality samples~\cite{Panico2021}).
In the Supplementary Information, we provide movies of simulated conservative (supplementary movie 1) and dissipative (supplementary movie 2) condensate evolution, cases discussed in the next sections~\cite{SImovies}.

\paragraph{\blue{Section III: Onsager condensation in the conservative case.}}
First, we focus on the investigation of clustering within the conservative case ($\gamma_\mathrm{LP} = 0$).
We compute dynamics for different initial inter-vortex lengths $\ell_\mathrm{v}^0$, which allow us to both i) gain an understanding of the vortex dynamics in terms of different system sizes; ii) reveal the interplay between the different mechanisms that regulate the clustering dynamics iii) use the conservative case as a reference for the dissipative cases analyzed in the next sections.

Throughout conservative dynamics, the total number of particles and angular momentum are constant in time. 
As a consequence, the healing length set at the beginning of the simulation will be preserved overall in the system dynamics $\xi(t) = \xi_0$.
%
%
\begin{figure}
    \centering
    \includegraphics[width=\columnwidth]{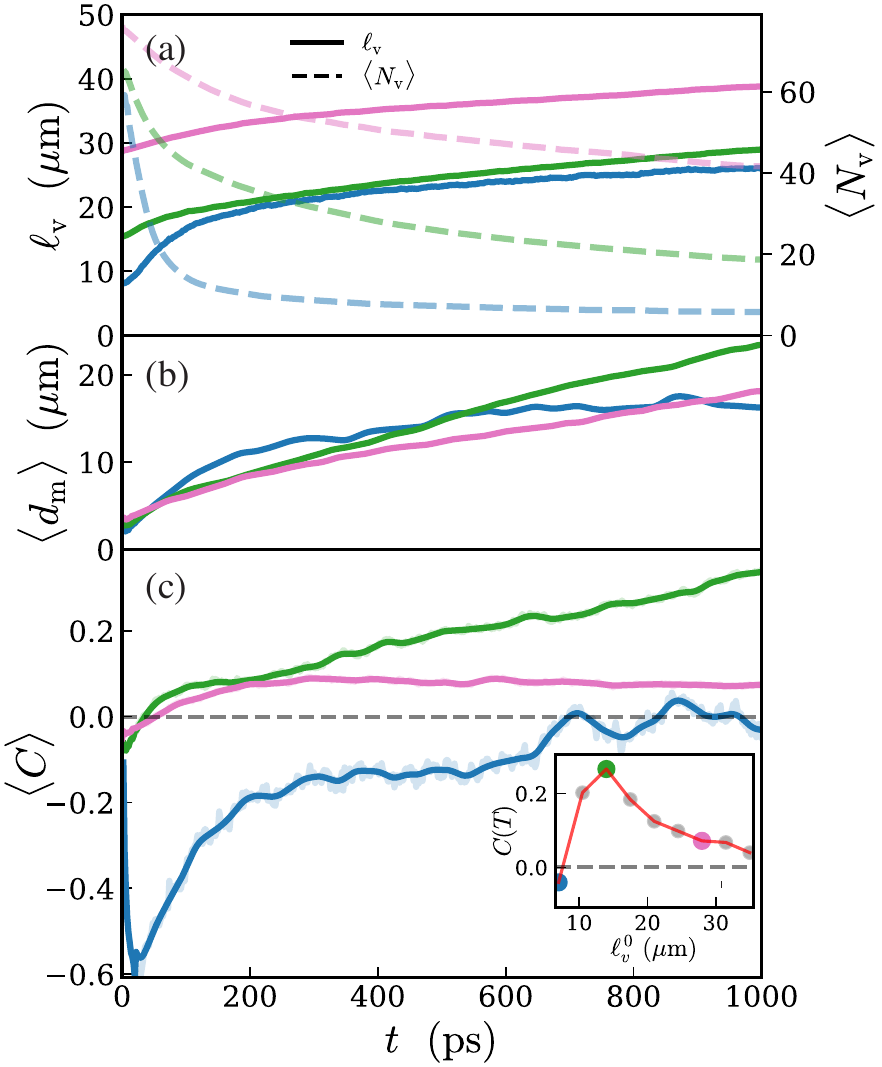}
    \caption{
    \textbf{Clustering in the conservative dynamics.} 
    \textbf{(a)} The average inter-vortex length $\ell_\mathrm{v}$ and total number of vortices (top), \textbf{(b)} averaged dipole moment $\langle d_\mathrm{m} \rangle$ (center) and \textbf{(c)} clustering correlation function $\langle C \rangle$ (bottom) are used to trace clustering of vortices along the system dynamics. The panels show the behavior of the different observables for different initial conditions of $\ell_\mathrm{v}^0$, at a fixed initial density (corresponding to healing length ${\xi_0 = 0.7 \mathrm{\mu m}}$). 
    The inset of the bottom panel shows the value of $\langle C \rangle$ calculated at a time $T=1$ ns. The interplay between early-time annihilation and long-time clustering mechanism leads to a maximum for $C(T)$ at ${{\ell_\mathrm{v}^0 = 14 \mathrm{\mu m}}}$.
    The faint colored data underneath the solid curves show the data averaged over $\mathcal{N}=10^2 $ realizations.
	}
	\label{fig2}
\end{figure}
In Fig.~\ref{fig2} we plot the relevant vortex and clustering observables along the system evolution: average number of vortices $\langle N_\mathrm{v} \rangle$ (panel a, dashed lines), inter-vortex length $\ell_\mathrm{v}$ (panel a, solid lines), average dipole moment $\langle d_m \rangle$ (panel c), and the clustering correlation function $\langle C \rangle$.
Ensemble averaging is applied over $\mathcal{N}=10^3$ different initial random vortex configurations. 
%
%
An example of vortex dynamics is given by the case depicted in pink color, corresponding to ${\ell_0 = 28 \mathrm{\mu m}}$. 
For this case, the real-space density distributions are reported in Fig.~\ref{fig1}, at different times: $t=0$, $0.5$ and $1$ $\mathrm{ns}$ (panels a, b, c, respectively).
The system evolves as follows: at the beginning, a large number of vortex-antivortex pair annihilate, followed by the consequent emission of sound waves.
In agreement with previous literature~\cite{simula2014prl,gauthier2019giant}, as the vortex cloud decays, evaporative heating processes take place followed by the tendency of the system to form same-sign vortex clusters of different sizes.
The decay towards a higher-temperature state is verified by the growth of the clustering quantities $\langle d_m \rangle$ and $\langle C \rangle$ towards larger positive values.
An example of vortex dynamics in the conservative case is provided in the supplementary movie 1~\cite{SImovies}. 

Let us now discuss how clustering behaves under the change of the inter-vortex length $\ell^0_\mathrm{v}$. 
Fig.~\ref{fig2}(a) shows that decreasing $\ell_\mathrm{v}^0$, the number of vortices exhibits a faster decay in time ---especially in the first transients of the evolution---; this shows that annihilation processes become more efficient as we increase the initial density of the vortex gas.
This behavior can be naively justified if one considers that the dynamics of a single interacting quantum vortex pair accelerates in the case two vortices are placed closer to one another~\cite{Barenghi_book}. 
Thus, a denser, therefore more strongly interacting, gas of defects (with opposite circulations in equal numbers), is expected to accelerate pair-collapse processes.
To tentatively quantify this behavior, we extract $\langle C \rangle$ at time  $T=1$ ns.
In the inset of Fig.~\ref{fig2}(c) we plot $C(T)$ as a function of the $\ell_\mathrm{v}^0$.
The data verifies the inverse proportionality between the two quantities, but up to a threshold value for $\ell_\mathrm{v}^0$ ($\sim  14 \mathrm{\mu m}$ for the set of parameters chosen) below which $\langle C \rangle^\mathrm{max}$ changes behavior and decreases.

Fig.~\ref{fig2}(c) shows that
this behavior can be explained by a quenching of the clustering processes taking place in the very early stage of dynamics  ($0 - 20 \mathrm{ps}$ for the case {$\ell_\mathrm{v}^0 \sim 8 \mu m$}, blue curves).
Inspection of the density profiles (see supplementary movie 1~\cite{SImovies}) reveals that the quenching effect is mainly originated by the fast vortex-antivortex annihilation events which, given the more dense gas of vortex-antivortex pairs, take place at shorter times-scales than the clustering processes.
As a result, in the diluted (therefore less interacting) vortex gas, the onset of clustering is {delayed}.
Only when the density of the vortex gas (due to the abrupt decay of the number of vortices) reaches sufficiently low values to slow down the vortex interactions and trigger the clusterization processes, the ``evaporative cooling" observed at early times transforms to ``evaporative heating", allowing $\langle C \rangle$ to grow towards positive values.
In the meantime, a large amount of incompressible energy has been already transformed to compressible energy, resulting in a total clusterization process being slowed down in comparison to cases with larger $\ell_\mathrm{v}^0$.
%

\paragraph{\blue{Section IV: Onsager condensation in the dissipative case.}}
\begin{figure}
	\centering
    \includegraphics[width=\columnwidth]{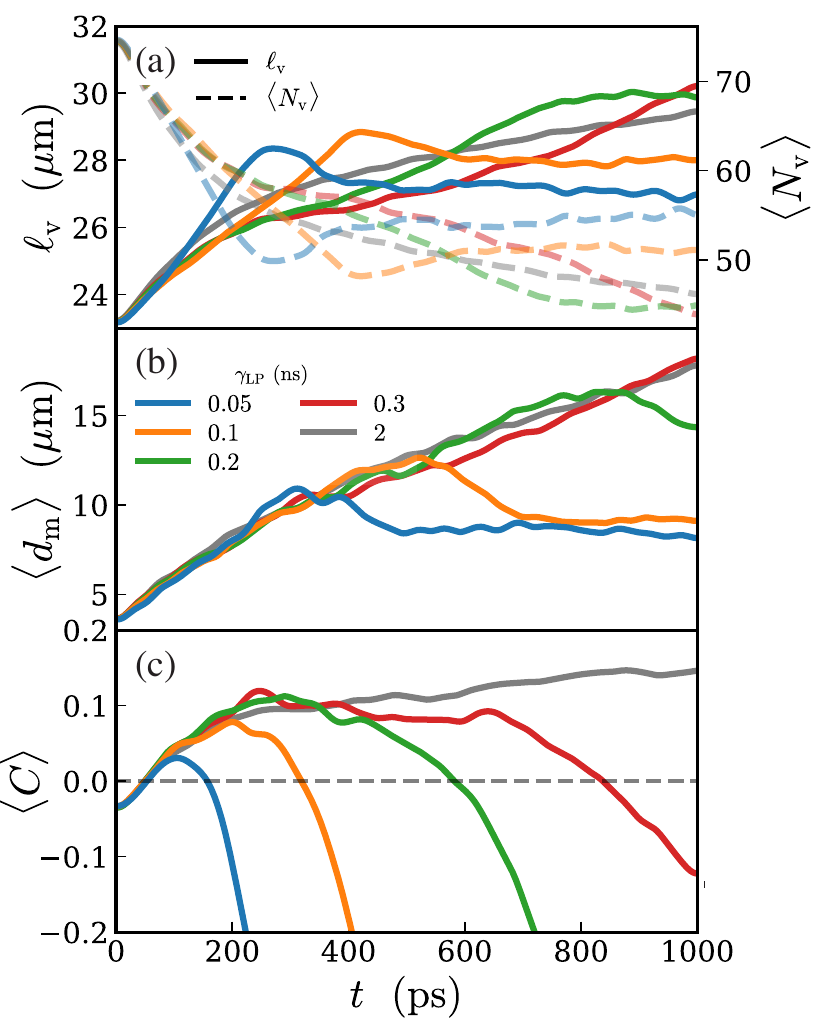}
	\caption{\textbf{Clustering in presence of dissipation}. The three quantities shown in Fig.~\ref{fig2} are calculated when turning on the dissipative term of Eq.~\eqref{eq:SGPE_pol}. 
    {The trap size is fixed in order that ${\ell_\mathrm{v}^0 = 28 \mathrm{\mu m}}$.
    For the system's parameters investigated, the curves show that the clustering mechanism is quenched as the polariton lifetime is decreased.}
    }
	\label{fig3}
\end{figure}

\begin{figure}[t]
	\centering
    \includegraphics[width=\columnwidth]{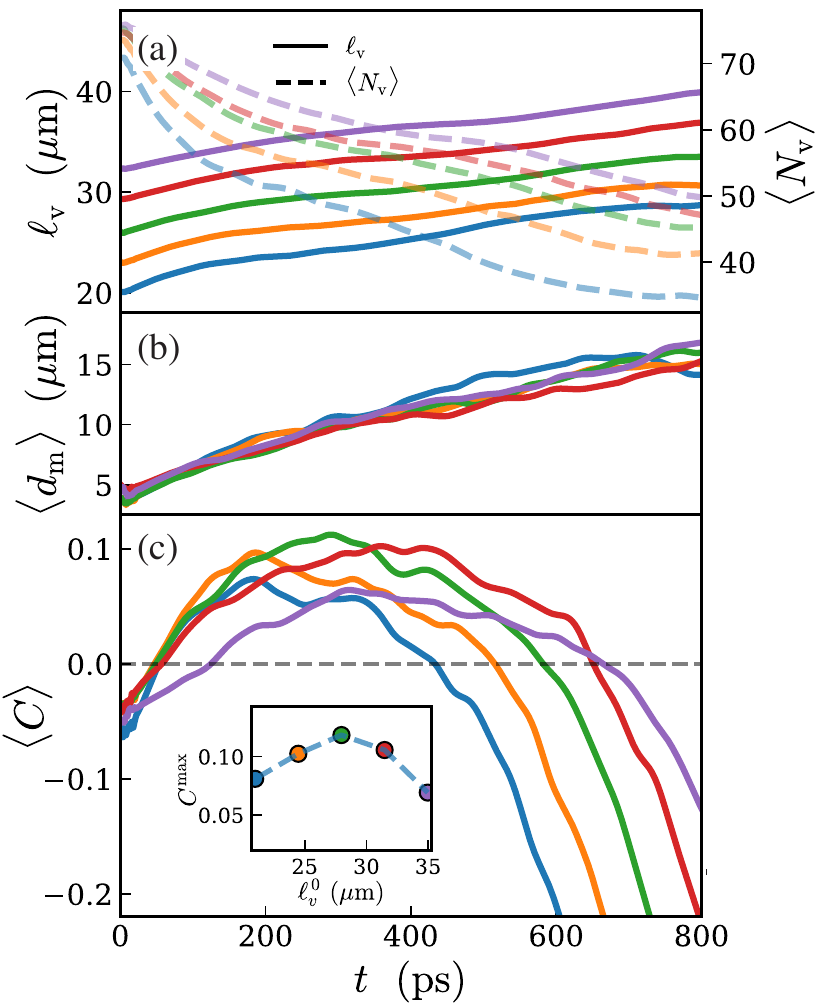}
	\caption{
     {\textbf{Clustering in the dissipative dynamics.}} 
    As in Fig.~\ref{fig2}, but for the dissipative case ($\gamma_\mathrm{LP} = 0.2 \mathrm{ns}$). 
    The green curve corresponds to the green curve in Fig.~\ref{fig3}.     The initial healing length reads ${\xi_0 = 0.7 \mathrm{\mu m}}$. %
    When the initial inter-vortex length $\ell_\mathrm{v}^0$ is varied, the interplay between the different dynamical mechanisms leads to a maximum in the clustering correlation function, found at ${\ell_\mathrm{v}^0 = 28 \mathrm{\mu m}}$, as shown in the inset of panel (c).
	}
	\label{fig4}
\end{figure}

We proceed by investigating how the vortex clustering is affected when accounting for a finite lifetime.
The main difference is that the dissipative Hamiltonian of the polariton equation of motion \eqref{eq:SGPE_pol} does not preserve the total average number of particles. 
Due to the cavity leaks (modeled by the term proportional to $\gamma_\mathrm{LP}$), the polariton density exponentially decays in time as $|\psi_\mathrm{LP}|^2~\propto~n_0~\exp{(-t/\tau_\mathrm{LP})}$.
Consequently, dissipation introduces a temporal dependence to the healing length \eqref{eq:heal_length}, which is redefined as $\xi(t) = {1}/{\sqrt{2 m g n_0\exp{(-t/\tau_\mathrm{LP})}}}$.
This is expected to bear important consequences on the vortex dynamics, and it is still unclear if (and in which range of realistic system parameters) vortex clustering occurs.

In Fig.~\ref{fig3} we re-propose the same analysis as in Fig.~\ref{fig2} for different values of the particle lifetime $\tau_\mathrm{LP}$, in a system with fixed geometry $D$ (i.e. fixed $\ell_\mathrm{v}^0$) and fixed initial density $n_0$.
We choose $\tau_\mathrm{LP}$ from the range of typical experimental polariton lifetimes $\in [0.05,2]$ $\mathrm{ns}$.
Similarly to the conservative case discussed in the previous section, clustering observables are monitored along the system evolution.
Importantly, Fig.~\ref{fig3}(c) shows that, despite the presence of dissipation, the clustering processes are still able to take place at early-stage dynamics, where $\langle C \rangle$ and dipole moment grow. 
We note, however, that the onset of clustering takes place only up to a certain time, after which $\langle C \rangle$ starts decreasing due to the effects of dissipation. 
This results in the presence of a maximum in the dynamical curve of the clustering correlation function.
Fig.~\ref{fig3}(c) shows that both i) the maximum value of $\langle C \rangle$ and ii) the times-scale measured along the clustering process increase as a function of $\tau_\mathrm{LP}$.

The condensate profiles depicted in Fig.~\ref{fig1}, panels (d), (e) and (f), show that this behavior is intrinsically linked to the time dependence of the size of a single vortex core.
An example of the vortex dynamics in the dissipative case is also provided in the supplementary movie 2~\cite{SImovies}. 
The density profiles show that as the density decays, the vortex cores expand, slowing down the vortex gas dynamics.
The slowing down of the vortex dynamics corresponds to a deceleration of the clustering of the vortex gas.
This is explained by a decrease of the energy per single vortex, which scales as $E_v \sim \ln{(1/\xi)}$: along the system dynamics, the incompressible energy ---stored in the vortex cloud--- transforms in compressible energy, compromising the evaporative heating.
We note that eventually, in the final part of the dynamics, the system reaches very low density, allowing small density fluctuations to generate new random vortex-antivortex pairs. This interferes further with the clustering processes. 
We conclude by noting that the clustered fraction of the vortex cloud $\langle C \rangle$ starts decreasing much earlier than dipole moment $d_\mathrm{m}$.

\begin{figure}[t]
	\centering
    \includegraphics[width=\columnwidth]{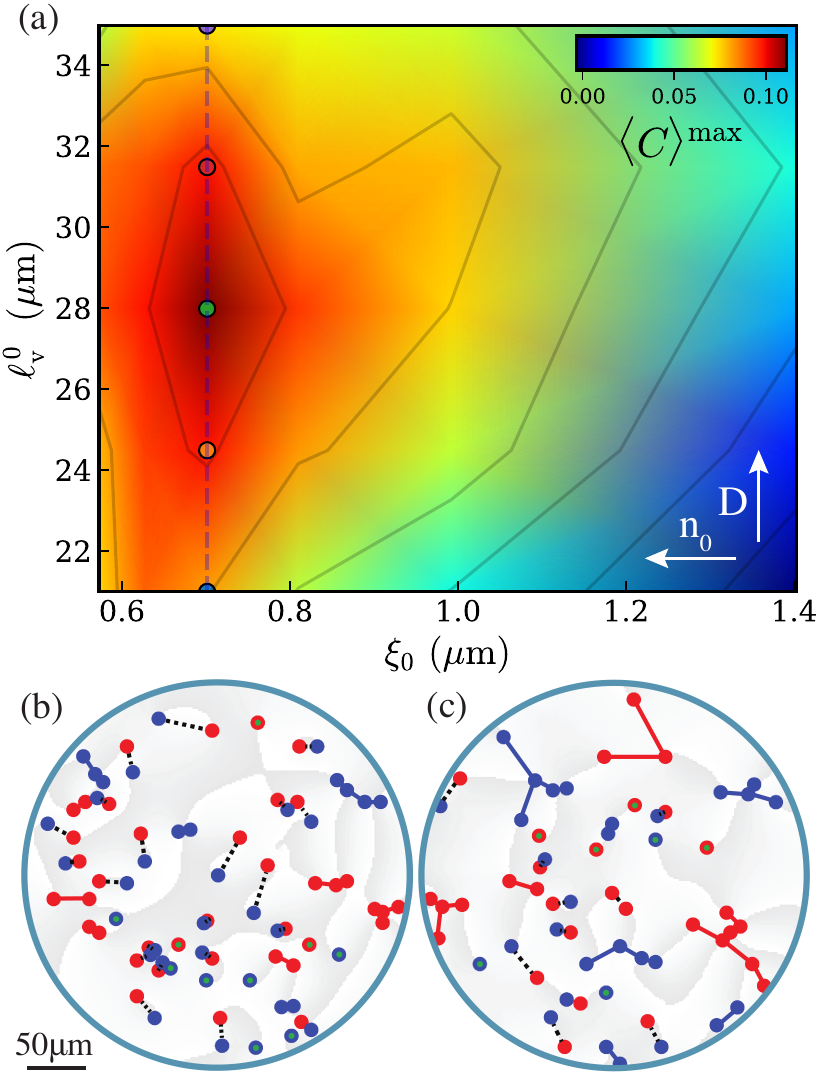}
	\caption{{\textbf{Clustering phase diagram}. \textbf{(a)} The maximum of the clustering correlation function ($\langle C \rangle^{submissio\mathrm{max}}$), calculated over $1 \mathrm{ns}$ of the polariton evolution, is numerically extracted and plotted as a function of the initial inter-vortex length $\ell_\mathrm{v}^0$ and initial healing length $\xi_0$.
    The maximum of the resulting two-dimensional surface, identified as a green dot, corresponds to the optimal condition for observing the onset of Onsager condensation (which also corresponds to the green curve in fig.~\ref{fig4}(c) and the green dot in the inset of Fig.~\ref{fig4}).
    }
    Single-realization snapshots of the phase are reported for  $t=0$ ps \textbf{(b)}, and for the optimal case at $t=300$ ps \textbf{(c)}. Positive (negative) circulation vortices are reported as blue (red) dots, and classified~\cite{johnstone2019evolution} into single vortices (dots with a green center), dipoles with opposite sign (dots connected with a black dashed line), and same-sign dipoles and clusters (dots connected with a line with the same color of the polarization).
}
	\label{fig5}
\end{figure}


\paragraph{\blue{Section V: Optimal condition for Onsager condensation.}}
We proceed to study the onset of Onsager condensation as a function of both the {confinement size $D$ (i.e. $\ell_\mathrm{v}^0$)}
and initial densities of the decaying system.
Given the competition of the different length- and time-scales of the relevant processes discussed in the previous sections, understanding the vortex dynamics and clustering becomes a non-trivial problem.

First, we solve the polariton dynamics for the dissipative case with $\gamma_\mathrm{LP} = 0.2 \mathrm{ns}$, with constant $n_0$ but varying the initial intra-vortex length: results are shown in fig.~\ref{fig4}.
The evolution of $\langle C \rangle$ depicted in Fig.~\ref{fig4}(c) shows that there exists an optimal curve which maximises the onset of vortex clustering. This is in analogy with the conservative case, but the optimal intervortex length is in general larger in the dissipative case.
Then, we extend our analysis by investigating the influence of the initial density $n_0$. 

Fig.~\ref{fig5}(a) depicts a carpet plot of $\langle C \rangle^\mathrm{max}$ as a function of the two relevant lengths of the system, namely $\ell_\mathrm{v}^0$ (proportional to the trap size $D$) and the initial healing length $\xi_0$ {(function of the initial density $n_0$, as in Eq.~\eqref{eq:heal_length} at $t=0$)}.
The vertical dashed blue line and points at $\xi_0= 0.7 \mathrm{\mu m}$ correspond to those depicted in the inset of fig.~\ref{fig4}(c).
The plot presents a maximum at $\xi_0= 0.7 \mathrm{\mu m}$ and {$l^0_\mathrm{v} = 28 \mathrm{\mu m}$}, which corresponds to the optimal condition for vortex clustering with experimentally available system parameters. 
The maximum is reached at $t = 300$ ps, as reported in fig.~\ref{fig4}(c); the real-space distribution of the vortex gas for the optimal case at the maximum point is shown, for a single realization, in fig.~\ref{fig5}(c), and compared with the initial vortex gas shown in fig.~\ref{fig5}(b).
Fig.~\ref{fig5}(a) also shows that for any fixed $\ell_\mathrm{v}^0$, by decreasing $\xi_0$ from large values, the clustering processes are enhanced. This is an expected behavior, as the energy per vortex would increase as its healing length decreases. 
Interestingly, we note that there exists an optimal condition also for $\xi_0$ {(namely $\xi_0 = 0.7 \mathrm{\mu m}$)}, below which $C$ decreases.
Investigation of the density and phase snapshots suggests this is due to the large amount of sound produced by the annihilation of vortex-antivortex pair ---produced in more quantity at larger densities---, sound which eventually hampers vortex clustering over the whole evolution.

\paragraph{\blue{Discussion and conclusions.}}
In this work, we give a qualitative and quantitative analysis of vortex clustering for dissipative condensates in the decaying turbulence scenario.
We study the evolution of an initially disordered vortex cloud towards more ordered states, revealing how the dynamics are affected by changing the relevant physical quantities, such as system size, initial density, and decay rates. 
Our investigation gives a clearer qualitative picture of the role and interplay of the different physical processes involved. 

The analysis of the conservative case, using realistic system parameters, already provides important information: while a denser and more interacting vortex cloud speeds up the clustering processes, there exists a threshold above which the annihilation prevails over clustering, diluting the gas too fast and retarding the evaporative heating mechanisms.
A similar scenario is also found modulating the initial density of the system: while larger densities speed up vortex clustering, there exists a threshold above which the sound emitted by the frequent vortex annihilations would slow down the motion of vortices.

As expected, when accounting for dissipation, evaporative heating is compromised. Our calculations reveal that, with respect to the conservative case, the optimal intervortex length is larger and the maximum value of the clustering correlation function is limited by the dissipation rate. However, we show that, for experimentally accessible parameters, the onset of vortex clustering can be observed in decaying polariton systems. Additionally, the persistence of a global dipole orientation over the nearest neighbor correlation $\langle C \rangle$ motivates further research and deeper insights on the mechanisms underlying the onset of turbulence in driven-dissipative condensates. In this context, the possibility to inject a large number of vortices and to precisely measure higher-order spatial correlations can be an asset of optical systems. Finally, these findings point towards the intriguing possibility of tuning the density and phase of exciton-polaritons quantum fluids to access different regimes of 2D quantum turbulence, ranging from superfluid turbulence to weak wave turbulence~\cite{berloff2010,koniakhin2020,Claude2020}. 
%

\paragraph{\blue{Acknowledgements. ---}}
We are grateful for stimulating discussions with A. S. Lanotte and D. Sanvitto. We acknowledge financial support from National Science Centre, Poland, grant no.~2016/22/E/ST3/00045. D.B and R.P. acknowledge financial support from PNRR MUR project PE0000023-NQSTI, financed by the European Union – Next Generation EU, and from EIC-Pathfinder project - ``Quantum Optical Networks based on Exciton-polaritons - Q-ONE" (Id: 101115575), financed by the European Union.

\end{document}